\documentclass[prlapplied,aps,twocolumn,longbibliography,showpacs,10pt]{revtex4-1}

\usepackage{graphicx}  
\graphicspath{{./Figures/}}
\usepackage{dcolumn}  
\usepackage{amssymb, amsmath}
\usepackage{hyperref}

\usepackage{graphicx}
\usepackage{bm}
\usepackage{color}

\definecolor{mygreen}{rgb}{0,0.5,0}
\definecolor{myred}{rgb}{0.75,0,0}
\definecolor{myblue}{rgb}{0,0,0.75}
\definecolor{mymagenta}{cmyk}{0,1,0,0.12}
\definecolor{mycyan}{cmyk}{1,0,0,0.12}
\definecolor{myorange}{rgb}{1,0.5,0}

\begin{document}

\title{Femtotesla direct magnetic gradiometer using a single multipass cell}
\author{V.\ G.\ Lucivero}\email[Corresponding author: ]{vito-giovanni.lucivero@icfo.eu}
\altaffiliation[Current affiliation: ]{ICFO-Institut de Ciencies Fotoniques, The Barcelona Institute of Science and Technology, 08860 Castelldefels (Barcelona), Spain}
\author{W.\ Lee}
\author{N.\ Dural}
\author{M.\ V.\ Romalis}
\affiliation{Department of Physics, Princeton University, Princeton, New Jersey, 08544, USA}
\date{\today}

\begin{abstract}
We describe a direct gradiometer using optical pumping with opposite circular polarization in two $^{87}$Rb atomic ensembles within a single multipass cell. A far-detuned probe laser undergoes a near-zero paramagnetic Faraday rotation due to the intrinsic subtraction of two contributions exceeding 3.5 rad from the highly-polarized ensembles. We develop analysis methods for the direct gradiometer signal and measure a gradiometer sensitivity of $10.1$ fT/cm$\sqrt{\mathrm{Hz}}$. We also demonstrate that our multipass design, in addition to increasing the optical depth, provides a fundamental advantage due to the significantly reduced effect of atomic diffusion on the spin noise time-correlation, in excellent agreement with theoretical estimate.
\end{abstract}

\maketitle

\twocolumngrid
\section{Introduction}
Operation of sensitive magnetic sensors in unshielded environment, including  Earth's magnetic  field and ambient noise, requires robust subtraction of common mode magnetic signals. Environmental noise suppression is a major challenge for several applications, such as non-invasive magnetoencephalography (MEG) \cite{Boto2018} and magnetocardiography (MCG) \cite{Jensen2016}
, magnetic anomaly detection \cite{Chu2017}, archeology \cite{David2004}, mineral exploration \cite{Nabighian2005} and search for unexploded ordnance \cite{NelsonJune2001}. The usual way of getting high noise cancellation and sensitivity improvement is to implement a magnetic gradiometer scheme. Highly-sensitive gradiometers are based on optically pumped magnetometers (OPMs), the most sensitive devices to measure low frequency magnetic fields to date \cite{Budker2007,Allred2002,Kominis2003,Dang2010}, and have been developed using either a single vapor cell with a multi-element photodiode in the spin-exchange-relaxation-free (SERF) regime \cite{Kominis2003,Dang2010} and at finite fields \cite{Smullin2009,Lucivero2019} or by using two microfabricated \cite{Sheng2017} or cm vapor cells \cite{Limes2020,Zhang2020} reaching sub-femtotesla sensitivity in multipass configuration \cite{Sheng2013}. Another promising approach is based on an actively shielded array of OPMs and it has been implemented in MEG \cite{Iivanainen2019}. Here we present a direct magnetic gradiometer that uses a single output with intrinsic subtraction of rotation signals from two atomic ensembles within one multipass cell. A similar approach has been used before with single pass configuration for rf magnetometry \cite{Wasilewski2010} and in a cw optical gradiometer \cite{Kamada2015}. Our multi-pass optical cavity design  using a 3-mirror ``V''-shape geometry has several advantages. It uses a single probe laser beam that passes repeatedly through two atomic ensembles that are polarized in opposite directions. As a result their optical rotation signals subtract, allowing for direct differential measurements. The direct cancellation of Faraday rotation from highly-polarized ensembles also avoids the complication of polarimeter signal wrap-around when the optical rotation exceeds $\pi/4$ radians in multipass geometry  \cite{Li2011}.  In addition, the optical design of the V-shaped multi-pass cell allows the laser beams to expand and overlap on multiple passes through the atomic ensembles. This reduces  diffusion broadening and increases correlation of spin measurements, unlike previous multi-pass cells that used distinct non-overlapping beams  which  were detrimental for possible sensitivity improvement by spin squeezing \cite{Sheng2013}. At the same time, the probe laser beam in the V-shaped multi-pass cell remains focused on one of the mirrors, which allows the laser to exit the cavity after a specific number of passes, in contrast to typical standing wave optical cavities \cite{Crepaz2015}. It also simplifies signal processing in the high-density and high-polarization regime where partial suppression of spin-exchange relaxation \cite{Appelt1999} causes highly non-linear spin evolution \cite{Sheng2013,Li2011}.\\
\begin{figure*}
\centering
	\includegraphics[width=\textwidth]{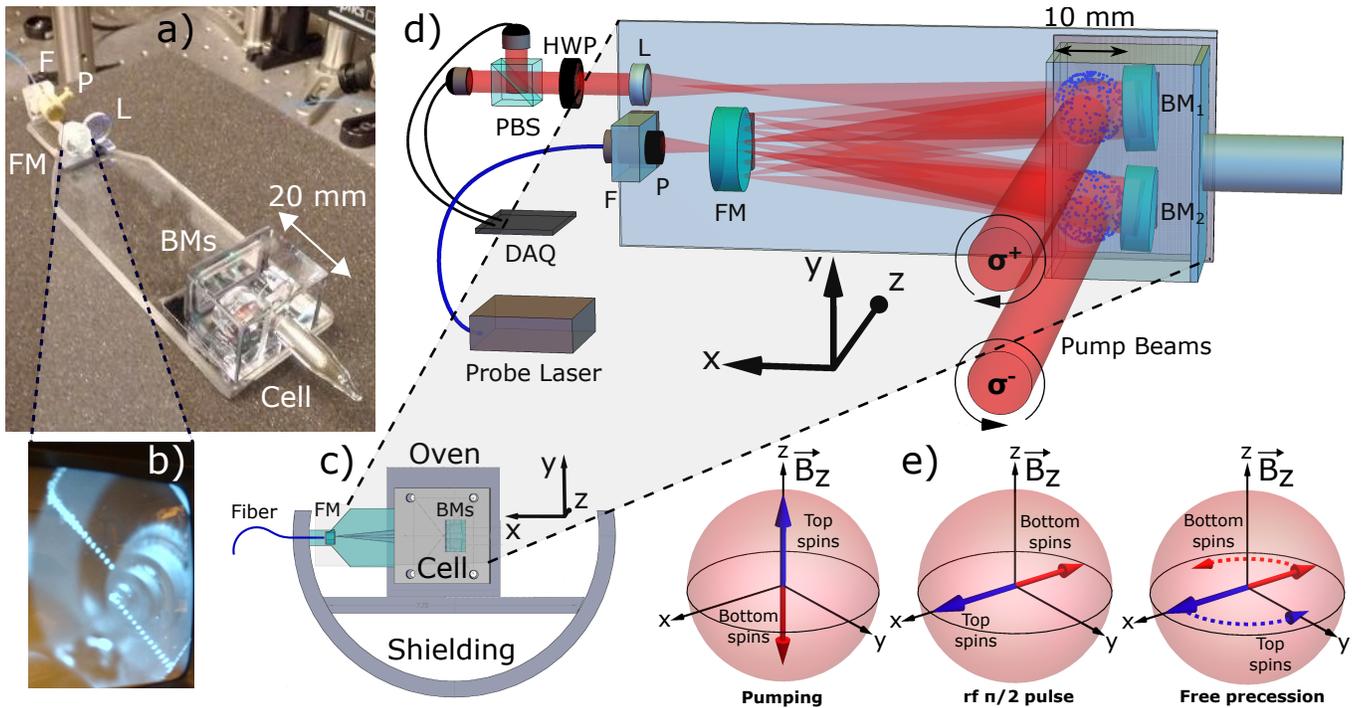}
	\caption{\textbf{a) V-Cell.} Picture of the sensor including probe input focuser (F) and polarizer (P), anodically-bonded front/back (FM/BMs) spherical mirrors and a $18\times20\times30$ mm$^3$ Pyrex cell enclosing both the back mirrors and the atomic vapor, output collimation lens (L). \textbf{b) Multipass geometry.} IR image (camera not shown) of the front mirror with probe beam spots after $60$ total passes through atomic ensembles before exiting the cavity. \textbf{c) Layout.} Cross-sectional view of the V-cell within a magnetic shielding with the Pyrex cell enclosed by a boron-nitride oven. A magnetic field gradient is applied in the y direction, while probe (input fiber coupled) and pump beams (free space) propagate in the x and z directions, respectively, as shown in d). \textbf{d) Full experimental sketch.} F, Focuser; P, Polarizer; FM, Front mirror; BM$_{1,2}$, Back mirrors; L, Collimation Lens; HWP, Half wave-plate; PBS, Polarizing beam splitter; DAQ, Data acquisition card. \textbf{e) Measurement sequence.} Optically-induced atomic orientation for top and bottom spins, angle tilt by $\pi/2$ pulse and free Larmor precession in the transverse plane.}
	\label{fig:setup}
\end{figure*}
Firstly, we describe the V-shaped sensor design, fabrication and working operation mode. Then, we report typical gradiometer signals with direct subtraction of Faraday rotation contributions with multiple wrapping and we introduce two analysis methods for sensitivity optimization in order to decouple the contribution of magnetic field gradient from amplitude and phase variations. We also study the fundamental spin quantum noise and we show that, by limiting atomic interaction to the region of overlapping beams with uniform large area, the spin noise spectrum has a nearly Lorentzian shape, whose linewidth is dominated by spin-exchange collisions rather than atomic diffusion, in contrast to prior work \cite{Sheng2013,Li2011}. We confirm this result by reporting a slower decay time of the experimental diffusion component of the spin time-correlation function relative to prior work \cite{Sheng2013}, in excellent agreement with theory \cite{Lucivero2017}.\\
\section{Sensor design and experimental setup}
The sensor, shown in Fig. (\ref{fig:setup}-a), consists of three half-inch convex spherical mirrors, with 100 mm radius of curvature, that are actively aligned in a V-geometry to give the desired beam propagation and then anodically bonded through silicon wafers to a Pyrex plate. The probe laser is fiber coupled, linearly polarized and focused at near-zero angle into a 170 $\mu$m hole made at the center of the front mirror. Then it expands to a beam diameter of 3.6 mm at the back mirrors where it nearly overlaps while undergoing 60 multiple reflections between front and back mirrors. Due to the non-zero input angle the probe is refocused to the front mirror at different spots, in number equal to half of total beam passes as shown in the infrared picture in Fig. (\ref{fig:setup}-b), before exiting the cavity. In order to make the atoms interact with a uniform wide beam, the atomic interaction is limited to the back region where a 2 cm wide Pyrex cell, which encloses the back mirrors, is also bonded to the plate through a second silicon wafer. The cell has an anti-reflection coated front window and is filled with pure $^{87}$Rb and $p_{N_2}=90$ Torr of $N_2$ buffer gas. Optical probe beam transmission through the cavity after 60 passes is typically greater than 50\%. The cell is heated with an ac electric current in a boron-nitride oven while the temperature is monitored by a thermocouple and stabilized to $0.1^\circ$C. The gradiometer structure stands within $5$ $\mu$-metal layers of magnetic shielding, as shown in Fig. (\ref{fig:setup}-c), and a concentric set of cylindrical coils (not shown). These generate the main field $B_z$ and a uniform gradient $\partial B_z/\partial y$. The experimental scheme is shown in Fig. (\ref{fig:setup}-d), in a simplified sketch with $12$ probe passes. We define two atomic interaction areas, addressed as top and bottom regions, where atoms are optically pumped with opposite polarization in the gradiometer operation mode with a baseline of $1.4$ cm. After multiple reflections the probe output is collimated and detected with a conventional balanced polarimeter, whose differential signal is fed into a digital oscilloscope.  The pump laser is a cw diode laser which is amplified by a tapered amplifier and tuned to the $^{87}$Rb $D_1$ line. A pulsed regime is generated by an acousto-optic-modulator and the first order diffracted beam is expanded and splitted in two parallel beams matching the atomic interaction areas. Top and bottom pump beams are circularly polarized with opposite ellipticity, $\sigma^+$ and $\sigma^-$, by two different quarter waveplates and are aligned along the z-axis to maximize initial atomic polarization. Atoms are pumped in the F=2 hyperfine state with $m_F=2$ and $m_F=-2$, i.e. parallel and anti-parallel with respect to the main field $B_z$, respectively. The measurement sequence is depicted in Fig. (\ref{fig:setup}-e). After $3$ ms of cw optical pumping, we apply a $\pi/2$ rf pulse with 30 cycles to the $B_y$ coil to flip the spins in the transverse plane. At this point top and bottom spins have opposite orientation in the x-direction, corresponding to a $\pi$ phase difference. We also apply an out-of-phase rf pulse to the gradient coil $\partial B_x/\partial y$ to create a small phase difference in spin precession signals for the two arms that compensates for the finite opening angle of the probe beam in the two arms of the V-cell. After the $\pi/2$ pulse, the spins freely precess at the Larmor frequency $\nu_L=(\gamma/2\pi)B_z$, where $\gamma=g_F\mu_B/\hbar$ is the gyromagnetic ratio, and we continuously record the free induction decay (FID) using paramagnetic Faraday rotation of the probe laser. The entire pump-tilt-probe cycle is repeated at driving period of $\tau=16.666$ ms.\par
\section{Experimental results and data analysis}
The FID output signal of the polarimeter is given by:
\begin{eqnarray}
\label{eq:signal}
V(t) &=& V_{\rm ver}(t)-V_{\rm hor}(t) = \\
=V_0&\sin&   \left\{
2 \phi_0^{\rm top}\sin \left[2\pi\left(\nu_L+\frac{\Delta\nu}{2}\right)t+d_0^{\rm top}\right]e^{-t/T_2^{\rm top}} \right.
\nonumber \\
&&\left.  -2 \phi_0^{\rm bot}\sin\left[2\pi\left(\nu_L-\frac{\Delta\nu}{2}\right)t+d_0^{\rm bot}\right] e^{-t/T_2^{\rm bot}} \right\}. \nonumber
\end{eqnarray}
where $V_0$ is the voltage corresponding to full probe intensity, $\phi_0^{\rm top}$ ($\phi_0^{\rm bot}$), $d_0^{\rm top}$ ($d_0^{\rm bot}$) and $T_2^{\rm top}$ ($T_2^{\rm bot}$) are the maximum rotation \cite{Li2011}, the phase and the transverse relaxation time of top (bottom) atomic ensemble \footnote{For high initial polarization, the decay can be non-exponential due to partial spin-exchange relaxation suppression \cite{Sheng2013}. Furthermore, Eq. \ref{eq:signal} is valid and has been tested at finite fields where nonlinear Zeeman effect is negligible.}, while $\Delta\nu$ is the difference in precession frequency due to the magnetic gradient.
\begin{figure}
	\centering
	\includegraphics[width=\columnwidth]{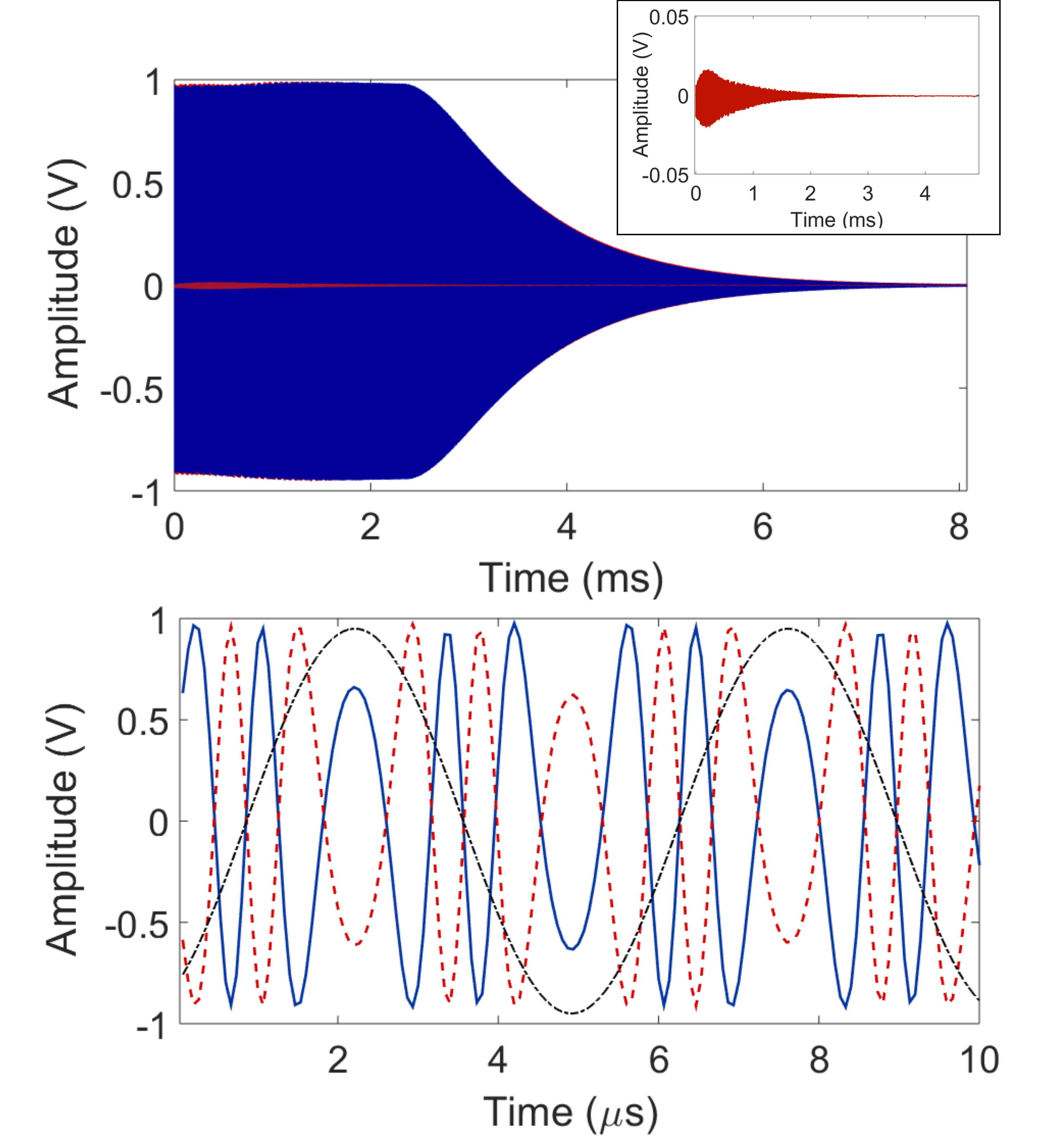}
	\caption{\textbf{Experimental signals. (Top)} Individual (blue) and differential (red) rotation signals at field $B_z=26$ $\mu$T, $400$ $\mu$W probe power and $T=120^\circ$C. \textbf{(Inset)} Zoom on the differential signal with direct cancellation at near-zero gradient. \textbf{(Bottom)} Individual signals over a shorter time scale showing a $\pi$ phase difference between the $V_{\rm top}$ (solid blue) and $V_{\rm bot}$ (dashed red) contributions when the atomic ensembles are polarized with opposite circular polarization. The signals undergo multiple zero-crossing over a Larmor period of spin precession (dot-dashed black).}
	\label{fig:signals}
\end{figure}
{In Fig. (\ref{fig:signals}) we report typical rotation signals $V_{\rm top}(t)$ and $V_{\rm bot}(t)$ at $T=120^\circ$C, obtained by blocking one of the two pump beams, respectively. Each contribution has a maximum rotation of about $3.5$ rad resulting in wrapping and multiple zero-crossing within a Larmor period. Over a shorter time scale, it is evident the opposite initial amplitude of the two signals. Then, when both pump beams are on, they directly subtract, as shown in the inset of Fig. (\ref{fig:signals}), resulting in an amplitude cancellation higher than $98\%$.
\begin{figure}
	\centering
	\includegraphics[width=\columnwidth]{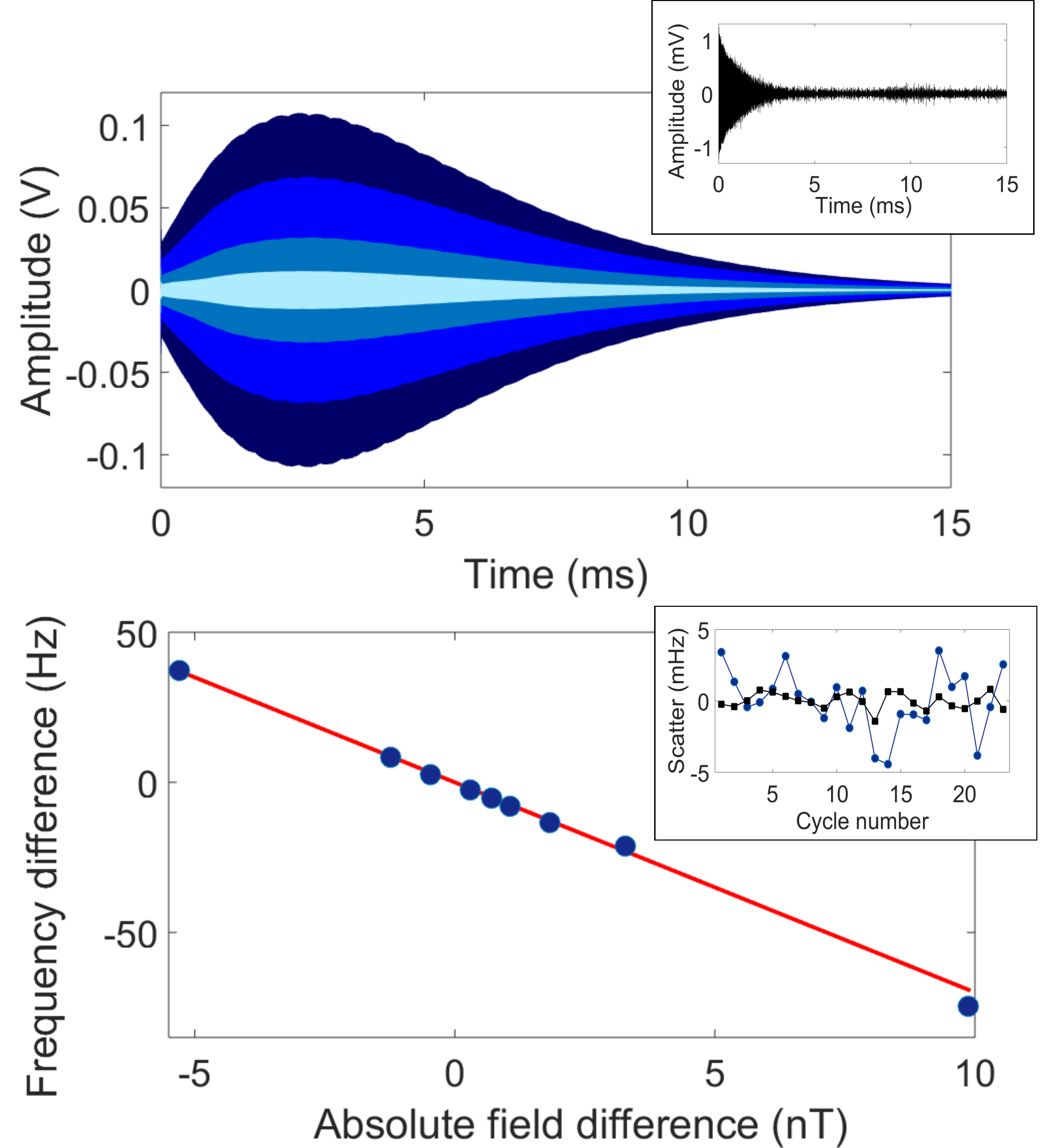}
	\caption{\textbf{(Top) Finite gradient gradiometer signal.} Gradiometer signal at applied gradient of (from top to bottom) $0.8$ nT/cm, $0.5$ nT/cm, $0.3$ nT/cm and $0.1$ nT/cm. \textbf{Inset.} Signal in absence of optical pumping showing fundamental noise and residual spin excitations. \textbf{(Bottom) Calibration.} Experimental (blue points) frequency difference and nominal gradient slope (red line) as a function of applied magnetic field difference. \textbf{Inset.} Frequency difference scatter over multiple measurements for polarized (blue dots) and unpolarized (black squares) atoms.}
	\label{fig:gradsignal}
\end{figure}
In Fig. (\ref{fig:gradsignal}) we show the dependence of this special direct signal on the applied uniform gradient at $T=100^\circ$C, optimized for sensitivity. Due to the difference in precession frequency, the signal builds-up from near-zero to reach a maximum, proportional to the applied gradient, then it decays due to relaxation in both contributions. We developed two complementary data analysis and optimization procedures. In the first strategy, we independently fit the two contributions to get probe voltage amplitude $V_0$, initial rotation amplitudes $\phi_0$, precession frequencies $\nu_L$, phases $\phi_0$ and relaxation times $T_2$. While these values typically agree within $1\%$ between top and bottom signals, in Eq. (\ref{eq:signal}) we replace  $\phi_0^{\rm bot}=\phi_0^{\rm top}+\Delta\phi$, $T_2^{\rm bot}=T_2^{\rm top}+\Delta T_2$, $d_0^{\rm bot}=d_0^{\rm top}+\Delta d_0$ and we perform a third fit to the direct gradiometer signal with \{$\Delta\nu$, $\Delta\phi$, $\Delta T_2$, $\Delta d_0$\} as free parameters, to take into account residual variations in all variables. Note that all four variables generate distinct differences in the signal shape and can be determined independently. The frequency difference output  $\Delta \nu$ is shown in Fig. (\ref{fig:gradsignal}) as a function of the absolute magnetic field difference $\Delta B=(\partial B_z/\partial y)\Delta y$ generated by the externally applied gradient $\partial B_z/\partial y$, including an offset of about $0.3$ nT/cm for zeroing the residual background. The experimental slope is in very good agreement with the nominal gradient coils calibration of $0.55$ nT/(cm mA) and the $\Delta y=1.4$ cm gradiometer baseline. At fixed gradient, we perform repetitive measurements to get standard deviation in the frequency difference estimation $\sigma_{\Delta\nu}$ and the differential magnetic sensitivity $B_{\Delta\nu}=(2\pi/\gamma)\sigma_{\Delta\nu}/\sqrt{\Delta f}$, in $\mathrm{T/\sqrt{Hz}}$ units, where $\Delta f=1/(2T_m)$ is the gradiometer bandwidth and $T_m=2$ ms is the fitting time for each signal, which is chosen for optimal sensitivity \cite{Gemmel2010}. In the inset of lower Fig. (\ref{fig:gradsignal}) we report an optimal experimental scatter with a $2.2$ mHz standard deviation, resulting in a measured sensitivity of 14.2 $\mathrm{fT/\sqrt{Hz}}$, corresponding to a gradiometer sensitivity of 10.1 $\mathrm{fT/cm\sqrt{Hz}}$ for a $1.4$ cm baseline. The signal obtained without the pump beam is shown in the inset of upper Fig. (\ref{fig:gradsignal}). It includes spin noise and RF spin excitation of a residual spin polarization created by a slight circular polarization of the probe laser.  Using the same fitting procedure on the residual spin excitation gives a standard deviation of $0.6$ mHz, corresponding to a sensitivity of 3.6 $\mathrm{fT/\sqrt{Hz}}$. The latter closely approaches the fundamental value of 2.7 $\mathrm{fT/\sqrt{Hz}}$ obtained by numerical simulations of the signal, given by Eq. (\ref{eq:signal}) with the addition of photon shot noise and atomic spin noise, independently measured as described in details in the next section.

We also implemented a second signal analysis  method that allows real-time measurements of the gradient signal and sensitivity. For small $\Delta \nu$, $\Delta \phi_0$, $\Delta d_0$ and  $\Delta T_2$ one can expand  Eq. (\ref{eq:signal}) to obtain
\begin{eqnarray}
V(t)&=&V_0  \left[ 2 \phi_0 (\Delta d_0 +2 \pi \Delta \nu t) \cos (2 \pi \nu_L+d_0)\right. \\
&+&\left.2( \Delta \phi_0+ \phi_0 \Delta T_2 t/T_2^2) \sin(2 \pi \nu_L+d_0)\right] e^{-t/T_2} \nonumber
\end{eqnarray}

One can see that the gradient signal $\Delta \nu$ appears  out-of-phase from the individual top and bottom signals. We perform an appropriately-phased FFT on the FID data and look separately at the real and imaginary Fourier components which each depend on pairs of unknown parameters \{$\Delta \phi_0$, $\Delta T_2$\} and \{$\Delta \nu$, $\Delta d_0$\}, respectively. This is done experimentally by performing FFT on each shot in real time. To optimally extract the signal $\Delta \nu$ we multiply FID by a custom window function before doing FFT.  Generally the optimal window function to maximize SNR in the presence of white noise is a matched filter equal to the envelope of the signal \cite{Spencer2010}, in our case $w(t)=(t/T_2) e^{-t/T_2}$. We find that this real-time method gives similar sensitivity to the non-linear fitting approach.
\begin{figure}
	\centering
	\includegraphics[width=1\columnwidth]{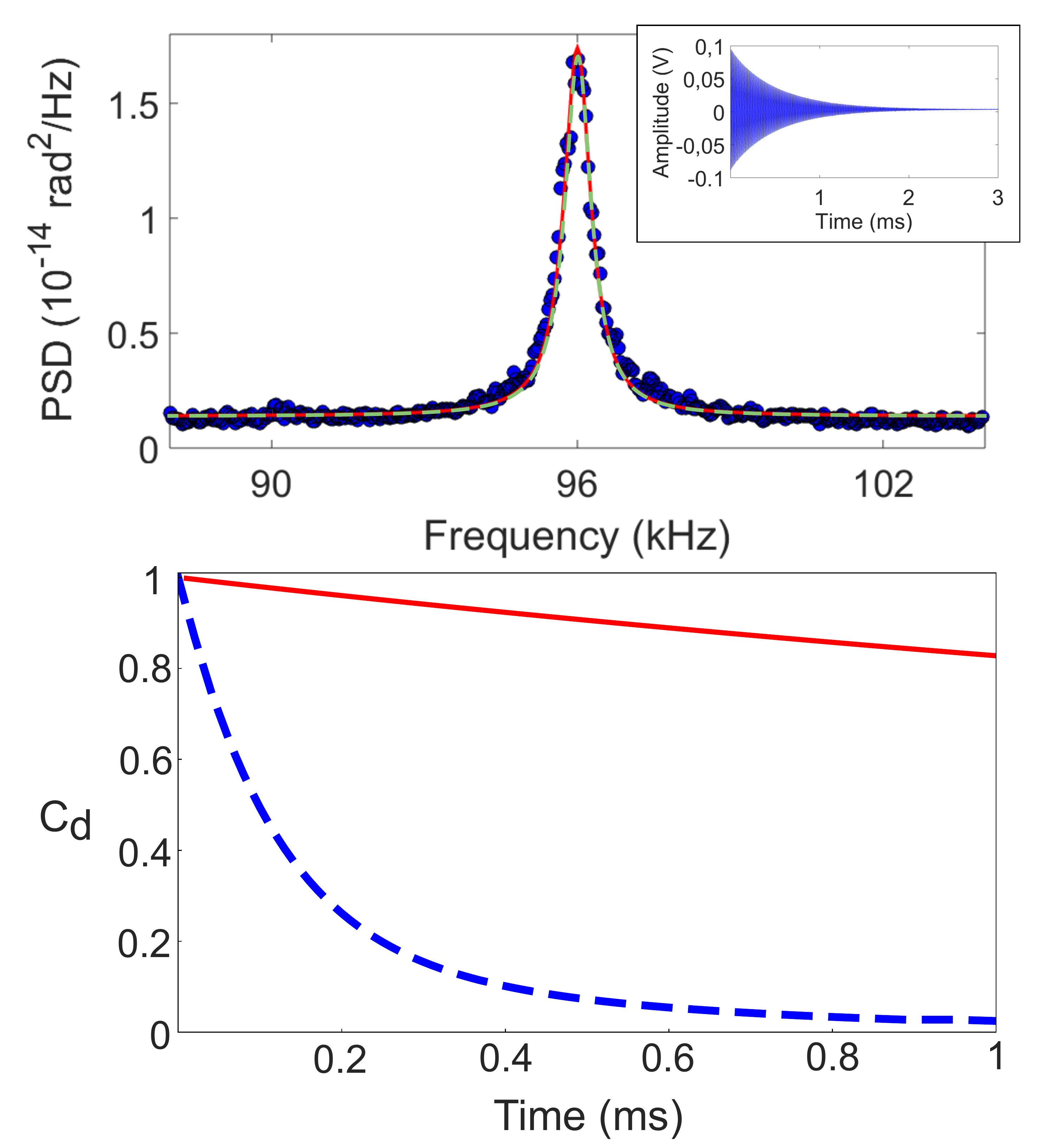}
	\caption{\textbf{(Top) Spin Noise Spectrum.} Experimental spin noise spectrum (blue points) centered at Larmor frequency $\nu_L=96$ kHz and a simple Lorentzian fit (red line) for unpolarized atomic ensemble under thermal equilibrium at $T=120$ C.  Prediction for spin noise spectrum including effects of diffusion without free parameters (green dashed line) \textbf{Inset.} Free induction decay signal for low initial polarization, where spin-exchange collisions limit the transverse relaxation time. \textbf{(Bottom) Comparison of diffusion correlation functions.} Calculated diffusion correlation function for the V-shaped multi-pass cell (red line) in comparison to the diffusion correlation function in cylindrical mirror multi-pass cavity\cite{Silver2005} used in \cite{Sheng2013}.}
	\label{fig:Figure4}
\end{figure}

\section{Noise analysis} When the atoms are polarized in the gradiometer operation mode, the sensitivity is limited by random spin excitations, due to rf broadband noise and pump fluctuations, resulting in a noise amplitude  higher than the fundamental atomic noise. In Fig. (\ref{fig:Figure4}) we report the spin noise power spectrum, measured at $T=120^\circ$C by probing intrinsic fluctuations of the unpolarized ensemble \cite{Zap1981} with a red-detuning of $200$ GHz from the $F=2$ state. We also don't apply the $\pi/2$ pulses, so the residual spin polarization of the atoms is not excited. One can see that the  peak spin noise power spectral density exceeds the background  noise power spectral density by more than a factor of 10, indicating good quantum-non-demolition (QND) resolution of the multi-pass cell.
The spin noise power spectrum is $S(\nu)=\langle\phi(t)^2\rangle\int_{-\infty}^{\infty}C(|\tau|)e^{-\imath 2\pi\nu\tau}d\tau$, where $C(|\tau|)=C_d(\tau) e^{-\tau/T_2}$ is the normalized spin noise time-correlation function, set by spin relaxation processes and the diffusion correlation function $C_d(\tau)$. The latter, which has been derived analytically in \cite{Lucivero2017} for arbitrary probe geometry, can modify the spectrum lineshape while it does not affect the total rotation noise variance $\langle\phi(t)^2\rangle$.
For $60$ passes through a 1 cm long Rb vapor with measured density $n=1.75\times10^{13}$ cm$^{-3}$  we calculate theoretical optical rotation r.m.s. noise of $\phi_{rms}^{th}=3.6\times10^{-6}$ rad \cite{Shah2010}. This is in good agreement with the measured area under the noise peak  $\phi_{rms}^{exp}=3.8\times10^{-6}$ rad after subtraction of the background noise floor of $\phi_{ph}=3.7\times10^{-8}$ rad/Hz$^{1/2}$. If the noise peak is fit to a Lorentzian (shown by a red line in Fig.  \ref{fig:Figure4}), it gives a width at half-maximum (FWHM) equal to $640$ Hz. The transverse relaxation time of a small coherent excitation obtained in the regime of low spin polarization, shown in the inset of Fig. (\ref{fig:Figure4}),  is equal to $T_2=0.55$ ms, which corresponds to a FWHM$=1/(\pi T_2)= 580$ Hz. The difference between the two linewidths is due to effects of atomic diffusion on the spin noise spectrum. However, due to the overlapping of multiple probe beams with uniform diameter of 3.6 mm the spectrum is nearly a pure Lorentzian shape, in contrast to prior work with multi-pass cells where, due to different transit times and varying probe focusing, the spin noise linewidth was limited by atomic diffusion, resulting in a distribution of Lorentzian functions \cite{Li2011,Sheng2013}. In the regime where the probe beam Rayleigh range is much larger than the length of the atomic vapor, the diffusion correlation function is given by $C_d^{th}(t)=1/(1+4tD/w_0^2)$ \cite{Lucivero2017}, where $D$ is the diffusion constant and $w_0$ is the Gaussian beam radius,  $D=1.5$ cm$^2$/s and $w_0=1.8$ mm in our cell.  The expected spin noise spectrum is given by Eq. (17) of \cite{Lucivero2017} and is shown with a green dashed line  in Fig. (\ref{fig:Figure4}) for  $T_2=0.55$ msec. One can see excellent agreement with the measured spectrum without any free parameters.
To illustrate the difference in the diffusion correlation functions, we show in Fig. (4) a comparison of the correlation function in the present experiment and in our prior multi-pass cavity design \cite{Sheng2013}.
The described multi-pass design, in addition to the high optical depth, provides a fundamental advantage with the possibility of improving the long-term sensitivity by spin squeezing \cite{Vasilakis2011}, because quantum correlations \cite{Wasilewski2010,Kong2020} could be preserved in a dense vapor despite of atomic diffusion.\\
\section{Conclusions}
We introduced a direct magnetic gradiometer showing a near-zero signal despite of the high optical rotation introduced by two atomic ensembles. The sensor consists of a single multipass cell, in contrast to prior geometries, based on either two vapor cells or two output signals \cite{perry2020alloptical}. The intrinsic cancellation of large polarization rotations, typical in state-of-the-art optical magnetometry, avoids complications related to signal processing. We developed two analysis methods for the special signal and we measured sensitivity of $10.1$ fT/cm$\sqrt{\mathrm{Hz}}$ with $1.4$ cm baseline and femtotesla projected sensitivity. The described gradiometer is also a multipass atomic sensor with a nearly pure Lorentzian spin noise spectrum, where atomic diffusion is not significantly affecting the time-correlation of the spin noise \cite{Sheng2013,Li2011}. In a quantum-noise-limited regime this would allow suppression of atomic spin noise due to spin squeezing \cite{Kong2020,Bao2020} while the sensitivity could be further improved by using a squeezed light probe \cite{Lucivero2016,Lucivero2017a}. While the described V-cell gradiometer has not yet been tested in unshielded environment, we expect that it will work particularly well for cancellation of broadband magnetic noise since it relies on direct real-time subtraction of two signals. We have successfully used an alternative approach based on two independent frequency measurements in  multi-pass cells for operation in unshielded environment  \cite{Limes2020}, but found it to suffer from reduced sensitivity in the presence of high-frequency magnetic noise. Therefore, a direct gradiometer is more promising for applications in challenging environments \cite{fu2020sensitive}. Finally, thanks to the anodic bonding fabrication technique, it could be further miniaturized \cite{Kitching2018}.\\
\section{Acknowledgements}
This work was supported by the DARPA AMBIIENT program.
\bibliography{INTRINSICGRAD_c}

\end{document}